\documentclass[letterpaper, 10 pt, conference]{ieeeconf}  

\IEEEoverridecommandlockouts                              

\pdfoutput=1 

\usepackage{cite}
\usepackage{amsmath,amssymb,amsfonts}
\usepackage{algorithmic}
\usepackage{graphicx}
\usepackage{textcomp}
\usepackage{xcolor}
\usepackage{graphicx}
\usepackage{amsmath}
\usepackage{amsfonts}
\usepackage{amssymb}
\usepackage[version=4]{mhchem}
\usepackage{siunitx}
\usepackage{bm}
\usepackage{algorithm}
\usepackage{algorithmic}
\usepackage{longtable,tabularx}
\usepackage{multirow}
\usepackage{subfigure}
\usepackage{caption}
\usepackage{placeins}
\usepackage{float}
\usepackage{nomencl}
\usepackage{hyperref}
\usepackage{inputenc}

\def\bf{\textbf}
\def\be{\begin{equation}}
\def\ee{\end{equation}}
\def\bes{\begin{equation*}}
\def\ees{\end{equation*}}
\def\ba{\begin{array}}
\def\ea{\end{array}}
\def\bba{\left[\begin{array}}
\def\eba{\end{array}\right]}
\def\bsa{\def\arraystretch{1.25}\left[\begin{array}}
\def\esa{\end{array}\right]
\def\arraystretch{1}}
\def\bea{\begin{eqnarray}}
\def\eea{\end{eqnarray}}
\def\beas{\begin{eqnarray*}}
\def\eeas{\end{eqnarray*}}
\def\beass{\begin{eqnarray**}}
\def\eeass{\end{eqnarray**}}

\newcommand{\TR}[1]{{#1}}
\newcommand{\YL}[1]{{#1}}

\title{\LARGE \bf
{Learning Based Model Predictive Control for Quadcopters with \\Dual Gaussian Process}}

\author{{Yuhan Liu and Roland T\'oth}
\thanks{{Y. Liu is with the Harbin Institute of Technology, Department of Control Science and Engineering, Harbin, China and the Eindhoven University of Tech., Dep.~of Electrical Eng., Control Systems Group, Eindhoven, The Netherlands. Email: y.liu11@tue.nl }}
\thanks{{R. T\'oth is with the Institute for Computer Science and Control, Budapest, Hungary and the Eindhoven University of Tech., Dep.~of Electrical Eng., Control Systems Group, Eindhoven, The Netherlands. Email: r.toth@tue.nl}}
\thanks{This research was supported by the Ministry of Innovation and Technology NRDI Office within the framework of the Autonomous Systems National Laboratory Program and China Scholarship Council (CSC).}
}

\begin{document}

\maketitle
\thispagestyle{empty}
\pagestyle{empty}

\begin{abstract}
\TR{An important} issue in quadcopter control is that \TR{an} accurate dynamic model \TR{of the system is} nonlinear, complex, and costly to obtain. \TR{This limits achievable control performance in practice.} Gaussian process (GP) \TR{based estimation} is an effective tool to learn \TR{unknown dynamics from} input/output data. \TR{However,} conventional GP-based control methods often ignore the computational cost \TR{associated with accumulating data during the operation of the system} and \TR{how to handle forgetting in continuous adaption}. In this paper, we present a novel Dual Gaussian Process (DGP) based model predictive control strategy that improves the performance of a quadcopter during trajectory tracking. The bio-inspired DGP structure is a combination of a long-term GP and a short-term GP, where the long-term GP is used to keep the learned knowledge in memory and the short-term GP is employed to \TR{rapidly} compensate \TR{unknown dynamics} during online operation. Furthermore, a novel recursive online update strategy for the short-term GP is proposed to successively improve the \TR{learnt} model during online operation in an efficient way. Numerical simulations are \TR{used to demonstrate} the effectiveness of the proposed strategy.

\end{abstract}

\section{INTRODUCTION}

Over the past decades, quadcopters have attracted \TR{significant} attention \TR{in} many research studies due to their wide \TR{applicability}\cite{agha2014health,hayat2016survey}. Flight control design for a quadcopter for agile maneuvering is not a simple task due to the quadcopter having six degrees of freedom with only four inputs to control the vehicle, which also leads to strong nonlinear couplings in the dynamics. Moreover, time-varying uncertainties and unmodeled dynamics, such as external wind, complex interactions of rotor airflows with the ground and \TR{objects such as walls}, friction and flapping dynamics of the rotor blades make the control design problem quite complex. \TR{Hence a} comprehensive and precise model of a quadcopter is still costly to obtain. Therefore, quadcopters \TR{mainly relay on} robust control \TR{solutions} against various disturbances and unmodeled dynamics, often limiting maneuvering capabilities to \TR{relatively} small pitch and roll angles. 

\TR{Recently, machine learning-based control has shown potential to save effort \TR{in} understanding unmodeled dynamics and achieving superior performance in compensatory control over physical model-based control. Especially, \emph{Gaussian Process} (GP) regression\cite{rasmussen2003gaussian}, which is a Bayesian nonparametric data-driven modeling method, is promising to automatically extract important features of the unknown dynamics from system input and output data. Gaussian process regression is a powerful nonparametric framework using  distribution over functions for nonlinear regression. The main advantage of GP regression is that it does not only provide a single function estimate, but also a variance, i.e. confidence interval, which indicates the accuracy of the regression result. In \cite{beckers2019stable}, a GP-based data-driven dynamic model was identified for reducing uncertainty of a given model and adapting the feedback gain by taking into account the model confidence. Furthermore, feedback linearization based control has been also achieved by learning a control affine GP model \cite{umlauft2017feedback}.}

In general, \YL{GP is not suitable for large training sets because the computational cost to train a GP is cubic in terms of the number of data points}. To solve this issue, various methods have been developed to reduce the computational load, but preserve the prediction accuracy. The simplest way is to choose a \emph{subset of data} (SoD) of size $M\ll N$ randomly to represent the full training set. However, this strategy can lead to serious underfitting as the probability of discarding crucial information included in the full training set grows with the reduction rate. \YL{In \cite{lawrence2003fast}, instead of choosing subset randomly, an \emph{information vector machine} based method is proposed to select the subset in a more intelligent way.} Besides the SoD method, many complicated techniques regarding more efficient model learning approaches have been explored, including \emph{Nystrøm method}\cite{williams2001using}, \emph{deterministic training conditional approximation} (DTC)\cite{csato2002gaussian,seeger2003bayesian}, \emph{partially independent training conditional approximation} (PITC)\cite{tresp2000generalized}, and \emph{fully independent training conditional approximation} (FITC)\cite{snelson2005sparse}. A unifying framework of the mentioned algorithms are discussed in \cite{quinonero2005unifying}. 
Recently, several studies for GP-MPC were carried out with the applications of race cars\cite{hewing2019cautious,kabzan2019learning}, robot arms\cite{carron2019data}, and quadcopters\cite{cao2016gaussian}.
However, most GP-based control methods do not perform online updating to avoid the issue of data bloating, although this would allow online adaptation with GP-based control during the system operation. By using data approximation techniques to counteract data bloating in online updating, the GP can gradually lose its knowledge learned during the training phase, and morph into a purely online adaptive regressor. Thus, designing a "memory"-based GP structure with online learning ability deserves further study. 

Inspired by the biological concept of "long/short term memory" of human beings, we propose a novel \emph{dual Gaussian process} (DGP) structure based model predictive controller for quadcopter trajectory tracking. The main contributions are: 1) By the authors' knowledge, the proposed DGP structure is the first framework that emphasizes both "memory" and "learning" ability by a combination of long/short-term GPs. As we show, such a method is capable of preventing forgetting relevant process information without the need of periodic reinforcement, but at the same time ensure exploration and adaptation to uncertain or varying  aspects of the dynamics. A long-term GP is utilized for the former while a short-term GP is used to for the latter objectives. 2) A novel recursive online updating strategy for the short-term GP is presented to continuously learn the unknown time-varying uncertainties during control operation. Numerical simulations are used to show the improvements regarding the performance of the baseline controller.

The paper is organized as follows. Section II introduces the mathematical model of the quadcopter and preliminaries of GP based regression and control. The novel recursive online sparse GP with dual structure is detailed in Section III. In Section IV, a DGP-MPC design procedure for quadcopters is presented. In Section V, simulation studies are provided to demonstrate the capabilities of the proposed method, while in Section VI conclusions on the achieved results are drawn.

\vspace{-0.5mm}
\section{Problem Formulation and Preliminaries}
\vspace{0.5mm}
\vspace{-1mm}\subsection{Problem Formulation} 

The quadcopter is modeled as a rigid body with four rotors. Let $\mathcal{I}$ denote the inertial frame and $\mathcal{B}$ denote the body fixed frame which is attached to the \emph{center of mass} (COM) of the quadcopter and oriented according to $\mathcal{I}$. Define $\bm{p}\in\mathbb{R}^{3}$ and $\bm{v}\in\mathbb{R}^{3}$ as the position and velocity of the COM.
\YL{Denoting $\bm{\zeta}=[\phi~\theta~\psi]^{\top}$ as the Euler angles of the quadcopter, 
then from the Newton-Euler formulation, one can obtain the system model for the quadcopter:
\be
\label{1}
\begin{aligned}
&\dot{\bm{p}}=\bm{v},  \quad m\dot{\bm{v}}=mg\bm{e}_{3}-T\bm{R}\bm{e}_{3}+\bm{F}_{\Delta},\\
&\dot{\bm{\zeta}}=\bm{\Theta}\bm{\omega},\quad\bm{J}\dot{\bm{\omega}}=-\bm{\omega}^{\times}\bm{J}\bm{\omega}+\bm{\tau}+\bm{\tau}_{\Delta}
\end{aligned}
\ee
where $\bm{R}$ is the rotation matrix from $\mathcal{B}$ to $\mathcal{I}$ with Z-Y-X sequence, $\bm{\omega}\in\mathbb{R}^3$ denotes the angular velocity of $\mathcal{B}$ with respect to $\mathcal{I}$, and the matrix $\bm{\Theta}\in\mathbb{R}^{3\times 3}$ is described as $\bm{\Theta}=\left[
c \theta ~  0  ~ \!-\!s \theta c \varphi ,
0~  1~  s \varphi , 
s \theta~  0 ~ c \theta c \varphi
\right]^{-1}$, $s$ and $c$ are short hands for sine and cosine, respectively.}
$m$ is the total mass of quadcopter, 
$\bm{e}_3=[0~0~1]^{\top}$ denotes the unit vector aligning with the gravity $g$ in $\mathcal{I}$, $[\cdot]^{\times}:\mathbb{R}^{3}\to\mathbb{R}^{3\times 3}$ denotes the cross-product operator, $\bm{J}\in\mathbb{R}^{3\times 3}$ is the inertia matrix of the quadcopter, $T\in\mathbb{R}^{+}$ is the thrust force generated by the four rotors, and $\bm{\tau}\in\mathbb{R}^{3}$ represents the control torque expressed in $\mathcal{B}$. The terms $\bm{F}_{\Delta}$ and $\bm{\tau}_{\Delta}$ represent unknown force and torque applied on the quadcopter resulting due to time-varying uncertainties and unmodeled dynamics, such as external wind, complex interactions of rotor airflows affected by the ground and walls, frictions and flapping dynamics of the rotor blades.

By denoting $\bm{x}=[\bm{p}^{\top}\quad\bm{v}^{\top}\quad\bm{\zeta}^{\top}\quad\bm{\omega}^{\top}]^{\top}\in\mathbb{R}^{12}$ and $\bm{u}=[T\quad\bm{\tau}^{\top}]^{\top}\in\mathbb{R}^{4}$, the continuous-time  model~\eqref{1} can be linearized and discretized resulting in the LTI model: \vspace{-0.5mm}
\be
\label{6}
\bm{x}(k+1)=\underbrace{\bm{A}\bm{x}(k)+\bm{B}\bm{u}(k)}_{\text{nominal}}+\underbrace{\bm{B}_d\bm{\Delta}(\bm{x}(k),\bm{u}(k))}_{\text{unknown}}
\ee
where $k\in\mathbb{Z}$ is the discrete time satisfying $t=kT_s$ with sampling time $T_s$ \textgreater $~0$, $\bm{A}\in\mathbb{R}^{12\times 12}$ and $\bm{B}\in\mathbb{R}^{12\times 4}$ are the known system matrices derived from the linearization of the idealistic rigid body dynamics, $\bm{\Delta}(\bm{x}(k),\bm{u}(k))=[\bm{F}_{\Delta}^{\top},\bm{\tau}_{\Delta}^{\top}]^{\top}\in\mathbb{R}^6$ denotes the unknown dynamics that are either resulted due to the approximation error of the applied linearization and discretization or they cannot be captured reliably by the idealistic rigid body model~\eqref{1} (i.e., both nonlinear effects and external disturbances), $\bm{B}_d\in\mathbb{R}^{12\times6}$ specifies that which states are influenced by $\bm{\Delta}$.  

\YL{In this paper, the control objective is to design an appropriate trajectory tracking controller for the system~\eqref{1} where the robust satisfaction of state and input constraints are guaranteed. In the simulation, the state constraint refers to a limited safe-space where the quadcopter could explore and the input constraint corresponding to the maximal thrust force and torque that the quadcopter could provide.}

\vspace{-1mm}\subsection{Learning with Gaussian Processes} 
To guarantee the aforementioned safety constraints, the unknown model $\bm{\Delta}(\bm{x}(k),\bm{u}(k))$ should be identified reliably. Our goal is to construct a probabilistic model of $\bm{\Delta}(\bm{x}(k),\bm{u}(k))$ from the system measurement data, and to improve accuracy gradually as more data are collected. We model  $\bm{\Delta}(\bm{x}(k),\bm{u}(k))$ as a GP where the state-input pairs $\bm{\tilde{x}}(k)\triangleq(\bm{x}^{\top}(k)\quad\bm{u}^{\top}(k))^{\top}\in\mathbb{R}^{12+4}$ and $\bm{\Delta}(\bm{\tilde{x}}(k)) \triangleq \bm{x}(k+1)-\bm{A}\bm{x}(k)-\bm{B}\bm{u}(k)$ are denoted as training inputs and outputs, respectively.  \YL{A GP is a distribution over functions, which is  full specified by:} 
\be
\label{7}
\begin{aligned}
&\bm{\mu}(\bm{\tilde{x}})= \mathbb{E}[\bm{\Delta}(\bm{\tilde{x}})],\\
&{k}(\bm{\tilde{x}},\bm{\tilde{x}}')= \mathbb{E}[(\bm{\Delta}(\bm{\tilde{x}})-\bm{\mu}(\bm{\tilde{x}}))(\bm{\Delta}(\bm{\tilde{x}}')-\bm{\mu}(\bm{\tilde{x}}'))],
\end{aligned} \vspace{-1mm}
\ee 
where $\bm{\mu}(\bm{\tilde{x}})$ is the mean function and ${k}(\bm{\tilde{x}},\bm{\tilde{x}}')\triangleq \mathrm{cov}(\bm{\Delta}(\bm{\tilde{x}}),\bm{\Delta}(\bm{\tilde{x}}'))$ is the positive semi-definite covariance function which denotes a measure for the correlation (or similarity) of any two data points $(\bm{\tilde{x}}, \bm{\tilde{x}}')$. Then we assume that: ${\bm{\Delta}}(\bm{\tilde{x}})\sim\mathcal{GP}(\bm{\mu}(\bm{\tilde{x}}),{k}(\bm{\tilde{x}},\bm{\tilde{x}}'))$.
The prior mean function is usually set to zero as no priori knowledge of $\bm{\Delta}(\bm{\tilde{x}})$ is given. Furthermore, there is a wide selection of the covariance functions, such as sinusoidal and Mat\'ern kernels, and in this paper we use the squared exponential function $k\left(\bm{\tilde{x}}, \bm{\tilde{x}}'\right)=\sigma_{f}^{2} \exp \left(-\frac{1}{2}(\bm{\tilde{x}}- \bm{\tilde{x}}')^{\top}\bm{\Lambda}^{-2}(\bm{\tilde{x}}- \bm{\tilde{x}}')\right)$ as a priori due to its universal approximation capability, where $\sigma_{f}^{2}\in\mathbb{R}_{\geq0}$ and $\bm{\Lambda}=\mathrm{diag}(\lambda_1,...,\lambda_d)$ are signal variance and length-scale hyperparameters, respectively.

Consider the training set $\mathcal{D}=\{\bm{\tilde{X}}=[\bm{\tilde{x}}_{1},...,\bm{\tilde{x}}_{N}],\bm{Y}=[\bm{y}_1,...,\bm{y}_N]\}$ with $N$ noisy output $\bm{y}_i=\bm{\Delta}(\bm{\tilde{x}}_i)+\bm{\epsilon}$ (i.e., corrupted by an $i.i.d.$ Gaussian noise $\bm{\epsilon}\sim\mathcal{N}(\bm{0} ,{\sigma}_{\epsilon}^{2}\bm{I})$ with standard deviation ${\sigma}_{\epsilon}$).
 The Gaussian prior distribution for the latent function $\bm{\Delta}$ and the model likelihood  of the training set $\mathcal{D}$ are given by $P(\bm{\Delta})=\mathcal{N}(\bm{\Delta}|\bm{0},\bm{K}_{N})$ and $P(\bm{Y}|\bm{\Delta}) = \mathcal{N}(\bm{Y}|\bm{\Delta},\sigma_{\epsilon}^2\bm{I}_{N})$ respectively, {where  $\bm{K}_{N}\in\mathbb{R}^{N\times N}$ denotes the kernel matrix with $[\bm{K}]_{ij}=k(\bm{\tilde{x}}_i,\bm{\tilde{x}}_j)$.}
Thus, according to the Bayes' rule, one can obtain the posterior distribution by \emph{maximum a posterior} (MAP) estimation\
$P(\bm{\Delta}|\bm{Y})\propto P(\bm{Y}|\bm{\Delta})P(\bm{\Delta}) 
\propto \mathcal{N}(\bm{\Delta}|\bm{K}_{N}(\bm{K}_{N}+\sigma_{\epsilon}^2\bm{I}_{N})^{-1}\bm{Y},(\bm{K}_{N}^{-1}+\sigma_{\epsilon}^{-2}\bm{I}_{N})^{-1})$. The associated posterior distribution of the function value $\bm{\Delta}({\bm{\tilde{x}}}^{*})$ at a new test point ${\bm{\tilde{x}}}^{*}$ can be derived by $P(\bm{\Delta}^{*}|\mathcal{D},\bm{\tilde{x}}^{*})=\int P(\bm{\Delta}^{*}|\bm{\tilde{X}},\bm{\Delta},\bm{\tilde{x}}^{*})P(\bm{\Delta}|\bm{Y})\mathrm{d}\bm{\Delta}
=\mathcal{N}(\bm{\mu}_{\Delta}(\bm{\tilde{x}}^{*}),\bm{\Sigma}_{\Delta}(\bm{\tilde{x}}^{*}))$
with the predictive mean and variance:
\be
\label{13}
\bm{\mu}_{\Delta}(\bm{\tilde{x}}^{*})=\bm{K}_{*N}(\bm{K}_{N}+\sigma_{\epsilon}^2\bm{I}_{N})^{-1}\bm{Y} 
\ee
\be
\label{14}
\bm{\Sigma}_{\Delta}(\bm{\tilde{x}}^{*})={k}_{**}-\bm{K}_{*N}(\bm{K}_{N}+\sigma_{\epsilon}^2\bm{I}_{N})^{-1}\bm{K}_{N*}
\ee
where $[\bm{K}_{N*}]_{j}={k}(\bm{\tilde{X}}_j,{\bm{\tilde{x}}}^{*})$, $\bm{K}_{*N}=\bm{K}_{N*}^{\top}$ and ${k}_{**}={k}({\bm{\tilde{x}}}^{*},{\bm{\tilde{x}}}^{*})$.
Furthermore, the marginal likelihood $P(\bm{Y})= \mathcal{N}(\bm{0},\bm{K}_{N}\!+\!\sigma_{\epsilon}^2\bm{I}_{N})$ and the hyperparameters $\bm{\theta}=[\sigma_{\epsilon}^{2},\sigma_{f}^{2},\lambda_1,...,\lambda_d]^{\top}$ can be optimized by maximizing
$\log P(\bm{Y})= -\frac{N}{2}\log 2\pi-\log|\bm{K}_{N}+\sigma_{\epsilon}^2\bm{I}_{N}|
-\frac{1}{2}\bm{Y}^{\top}(\bm{K}_{N}+\sigma_{\epsilon}^2\bm{I}_{N})^{-1}\bm{Y}$
by means of conjugate gradient-based algorithm\cite{rasmussen2003gaussian}, which has a computational load of $\mathcal{O}(N^3)$ per iteration.



\vspace{-0.5mm}
\section{Recursive Online Sparse Gaussian Process with Dual Structure}
\vspace{-0.5mm}\subsection{Sparse Gaussian Process}

The computational load per test case for the full  GP in~\eqref{13} and~\eqref{14} is $\mathcal{O}(N)$ and $\mathcal{O}(N^2)$ in terms of mean and variance, respectively. In this paper, we focus on the generalization of \emph{sparse variational GP regression} (SVGP), which has been first introduced in \cite{titsias2009variational}. 

The goal of sparse GP is to find a set of pseudo inputs $\bm{\tilde{X}}_\mathrm{u}\!=\![\bm{\tilde{x}}_{\mathrm{u},1},...,\bm{\tilde{x}}_{\mathrm{u},M}]$ corresponding to pseudo outputs $\bm{\Delta}_\mathrm{u}\!=\![\bm{\Delta}_{\mathrm{u},1},...,\bm{\Delta}_{\mathrm{u},M}]$ of size $M\ll N$. \YL{In SVGP, the true posterior distribution $P(\bm{\Delta}|\mathcal{D})$ is approximated by a Gaussian distribution $q(\bm{\Delta},\bm{\Delta}_\mathrm{u})\!=\!P(\bm{\Delta}|\bm{\Delta}_\mathrm{u})q(\bm{\Delta}_\mathrm{u})$ by means of variational inference, where a tractable Gaussian $q\left(\bm{\Delta}_\mathrm{u}\right)\!=\!\mathcal{N}(\bm{\Delta}_\mathrm{u}|\bm{m}_u,\bm{S}_u)$ is used,
corresponding to the maximization of the \emph{evidence lower bound} (ELBO) of $\log P(\bm{Y})$:
\be
\label{17}
\begin{aligned}
\mathcal{L}(q) &\triangleq \int q\left(\bm{\Delta},\bm{\Delta}_\mathrm{u}\right) \log \frac{P\left(\bm{Y}, \bm{\Delta},\bm{\Delta}_\mathrm{u}\right)}{q\left(\bm{\Delta},\bm{\Delta}_\mathrm{u}\right)} \mathrm{d} \bm{\Delta} \mathrm{d} \bm{\Delta}_\mathrm{u}\\
&=F(q)-\mathcal{KL}(q\left(\bm{\Delta}_\mathrm{u}\right)||P\left(\bm{\Delta}_\mathrm{u}\right)),
\end{aligned}
\ee
where $F(q)=\int q\left(\bm{\Delta}_\mathrm{u}\right)P\left(\bm{\Delta}|\bm{\Delta}_\mathrm{u}\right) \log P(\bm{Y} | \bm{\Delta}) \mathrm{d} \bm{\Delta}\mathrm{d} \bm{\Delta}_\mathrm{u}$, and $\mathcal{KL}(\cdot||\cdot)$ represents the Kullback-Leibler divergence which quantifies the similarity between two distributions. Then the optimal $q^{*}\left(\bm{\Delta}_\mathrm{u}\right)$ in terms of minimum of~\eqref{17} with mean $\bm{m}_u$ and variance $\bm{S}_u$ can be derived as:}
\be
\label{18}
\ba{l}
\bm{m}_{u}=\sigma_{\epsilon}^{-2}\bm{S}_{u} \bm{K}_{M}^{-1} \bm{K}_{M N}  \bm{Y}\\
\bm{S}_{u}=\bm{K}_{M}\left(\bm{K}_{M}+\sigma_{\epsilon}^{-2} \bm{K}_{M N} \bm{K}_{N M}\right)^{-1} \bm{K}_{M}
\ea
\ee
where $[\bm{K}_{MN}]_{ij}={k}(\bm{\tilde{x}}_{ui},{\bm{\tilde{x}}_{j}})$ denotes the covariance matrix between pseudo inputs $\bm{\tilde{X}}_u$ and training inputs $\bm{\tilde{X}}$, $\bm{K}_{MN} = \bm{K}_{MN}^{\top}$, $\bm{K}_{M}$ is the covariance matrix of pseudo inputs. Merging the optimal distribution  $q^{*}\left(\bm{\Delta}_\mathrm{u}\right)$ into $\mathcal{L}(q)$, we get: \vspace{-1.5mm}
\be
\label{19}
\mathcal{L}(q) = \log \mathcal{N}(\bm{Y}|\bm{0},\bm{Q}_{N}+\sigma_{\epsilon}^{2}\bm{I}_{N})-\frac{1}{2\sigma_{\epsilon}^{2}}\mathrm{tr}(\bm{K}_{N}-\bm{Q}_{N})\vspace{-1mm}
\ee 
where $\bm{Q}_{N}=\bm{K}_{NM}\bm{K}_{M}^{-1}\bm{K}_{MN}$. The pseudo inputs $\bm{\tilde{X}}_\mathrm{u}$ and hyperparameters can be optimized by maximizing~\eqref{19}, which results in a computational load of $\mathcal{O}(NM^{2}+M^{3})$. 

With the approximated posterior $q(\bm{\Delta},\bm{\Delta}_\mathrm{u})$, the marginal distribution of $\bm{\Delta}$ is $q(\bm{\Delta})=\int P\left(\bm{\Delta}|\bm{\Delta}_\mathrm{u}\right) q\left(\bm{\Delta}_\mathrm{u}\right)\mathrm{d} \bm{\Delta}_\mathrm{u}=\mathcal{N}(\bm{\mu}_{\Delta}(\bm{\tilde{x}}),\bm{\Sigma}_{\Delta}(\bm{\tilde{x}}))$
with mean and variance:
{\setlength\abovedisplayskip{3pt}
 \setlength\belowdisplayskip{1pt}
\be
\label{20}
\bm{\mu}_{\Delta}(\bm{\tilde{x}}^{*})=\bm{K}_{*M} \bm{K}_{M}^{-1}\bm{m}_{u}
\ee
\be
\label{21}
\bm{\Sigma}_{\Delta}(\bm{\tilde{x}}^{*})=k(\bm{\tilde{x}}^{*},\bm{\tilde{x}}^{*})\!-\!\bm{K}_{*M}\left(\bm{K}_{M}^{-1}\!-\!\bm{K}_{M}^{-1}\bm{S}_{u}\bm{K}_{M}^{-1}\right) \bm{K}_{M*} \vspace{-2mm}
\ee
}
where $[\bm{K}_{*M}]_{j}={k}(\bm{\tilde{x}}^{*},\bm{\tilde{x}}_{\mathrm{u},j})$.

It is worth mentioning that, the hyperparameters and pseudo points are fixed after the ELBO maximization, and the GP is used to predict the mean and variance at a new data point during the operation. However, due to the fact that the flying environment for the quadcopter is complicated and time-varying, the offline pre-trained GP model is required to be successively improved to ensure continuous adaptation.  

\vspace{-1mm}\subsection{Recursive Online Sparse Gaussian Process }

Next a novel online update strategy for sparse Gaussian processes is proposed. The proposed method updates the posterior mean and variance of $q\left(\bm{\Delta}_\mathrm{u}\right)$ recursively based on the regression error at the current time step $k$.

On the basis of recursive Bayesian regression, and given a new online measurement output $\bm{y}_k=\bm{\Delta}(\tilde{\bm{x}}_k)+\bm{\epsilon}$, the posterior mean and variance~\eqref{18} at the $k^\mathrm{th}$-step $q_k(\bm{\Delta}_\mathrm{u})=\mathcal{N}(\bm{\Delta}_\mathrm{u}|\bm{m}_{u}^{k},\bm{S}_{u}^{k})$ can be rewritten in terms of the posterior $q_{1:k-1}(\bm{\Delta}_\mathrm{u})=\mathcal{N}(\bm{\Delta}_\mathrm{u}|\bm{m}_{u}^{k-1},\bm{S}_{u}^{k-1})$:
\be
\label{22}
\begin{aligned}
&\bm{m}_{u}^{k} = \bm{S}_{u}^{k}(\bm{S}_{u}^{k-1}\bm{m}_{u}^{k}+\sigma_{\epsilon}^{-2}\bm{\Phi}_k^{\top}\bm{y}_k)\\
&\bm{S}_{u}^{k} = (\bm{S}_{u}^{k-1}+\sigma_{\epsilon}^{-2}\bm{\Phi}_k^{\top}\bm{\Phi}_k)^{-1}
\end{aligned}
\ee
with kernel $\bm{\Phi}_k=\bm{K}_{xM} \bm{K}_{M}^{-1}|_{\tilde{\bm{x}}=\tilde{\bm{x}}_k}$ and $[\bm{\Phi}]_{j} = \phi(\tilde{\bm{x}},\tilde{\bm{x}}_{uj})$. Then~\eqref{20} can be seen as a linear combination of $M$ kernel functions, each one is corresponding to a pseudo input:
{\setlength\abovedisplayskip{1pt}
 \setlength\belowdisplayskip{1pt}
\be
\label{23}
\bm{\mu}_{\Delta}(\bm{\tilde{x}})=\sum_{j=1}^{M}{m}_{uj}\phi(\tilde{\bm{x}},\tilde{\bm{x}}_{uj})= \bm{\Phi} \bm{m}_u
\ee
}
Note that~\eqref{23} can be interpreted as a weight-space representation of~\eqref{20}. Thus, recursive least squares constitutes an efficient way to update~\eqref{20} and~\eqref{21} with online streaming data points. Given a new data point $\{\tilde{\bm{x}}_k,  \bm{y}_k\}$, the posterior mean and variance can be updated by:
\be
\label{24}
\left\{
\ba{cl}

\bm{m}_{u}^{k} &=\bm{m}_{u}^{k-1}+\bm{L}_k\bm{r}_k\\

\bm{r}_k &= \bm{y}_k-\bm{\Phi}_k\bm{m}_{u}^{k-1}\\

\bm{L}_{k} &= \bm{S}_{u}^{k-1}\bm{\Phi}_k^{\top}\bm{G}_{k}^{-1}\\

\bm{G}_{k} &= \lambda + \bm{\Phi}_k\bm{S}_{u}^{k-1}\bm{\Phi}_k^{\top}\\

\bm{S}_{u}^{k} &= \lambda^{-1}(\bm{S}_{u}^{k-1}-\bm{L}_{k}\bm{G}_{k}\bm{L}_{k}^{\top})
\ea\right.
\ee
where $\lambda$ is the forgetting factor. The recursion starts from $\bm{m}_{u}^{0}$ and $\bm{S}_{u}^{0}$, which can be seen as the the prior of the GP. This recursive online sparse GP can be embedded into the predictor of an MPC scheme, and is capable of dealing with varying disturbances and sudden changes in the  dynamics. 

\vspace{-1mm}\subsection{Dual Gaussian Process Structure}
\begin{figure}[t]
  \captionsetup{font={small}}
  \includegraphics[width= 3.4in]{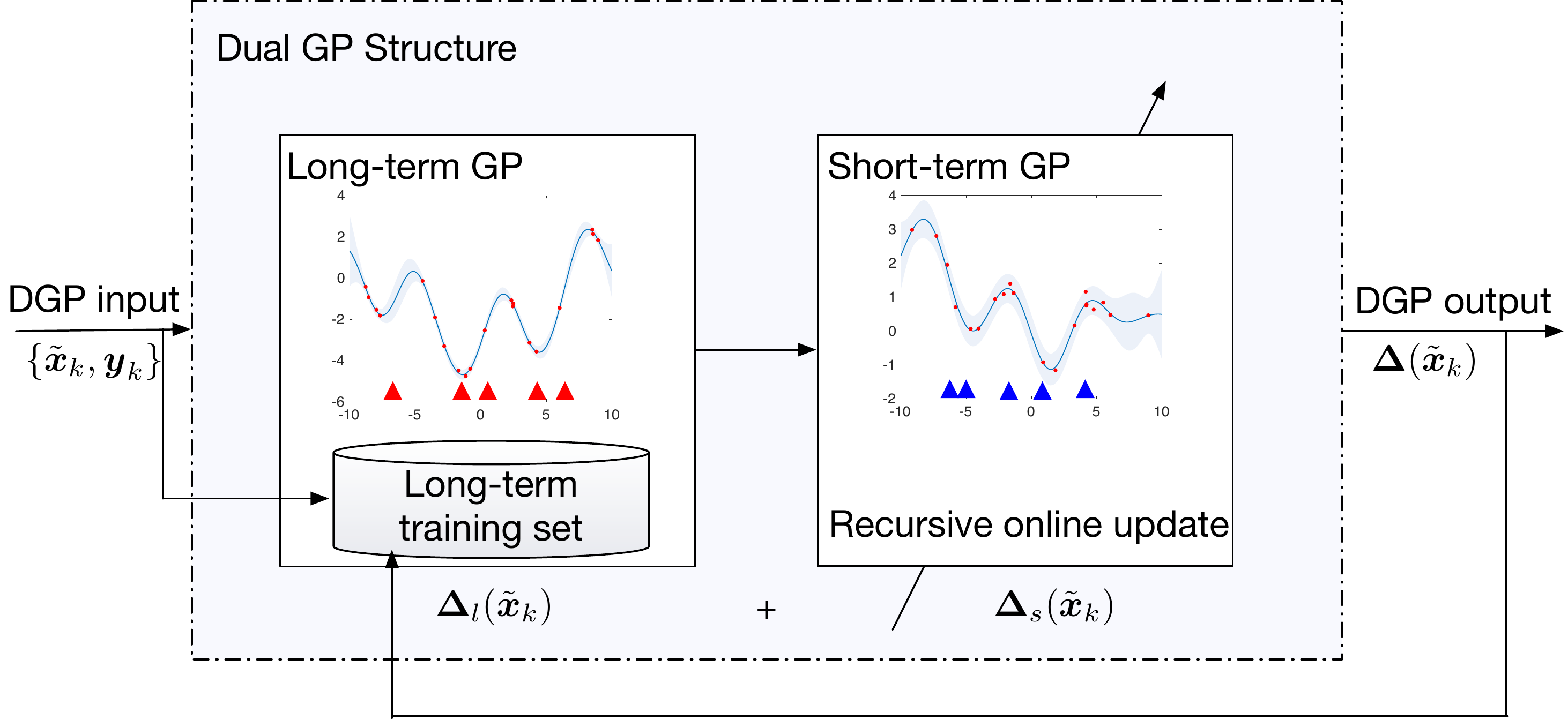}
  \caption{An illustration of the Dual Gaussian Process structure} 
  \label{fig:2}\vspace{-3mm}
\end{figure}

Despite the adapting capability of the recursive online sparse GP, the learned dynamics obtained during the offline training phase, will be forgotten during the online learning phase. This means that, if the current evidence during the online control phase does not support the dynamics seen, they will gradually disappear via the forgetting factor $\lambda$  according to~\eqref{24}. For example, the "knowledge" for the uncertainties with large roll and pitch angles will fade in case the quadcopter is hovering for a while.  Inspired by the biological concept of "long/short term memory", we present a novel structure named \emph{Dual Gaussian Process} (see Fig.~\ref{fig:2}), which allows the storage of collected experience and is able to prevent knowledge forgetting. 

The DGP structure mainly consists of two GPs: The first GP corresponds to the long-term GP, which is employed to learn all the time-independent uncertainties and disturbances, including aerodynamics besides the rigid body dynamics and indoor effects. The hyperparameters and the pseudo points of the long-term GP keeps fixed during online learning phase, and will be batch updated after each mission. This makes the long-term GP to have the ability to keep the offline trained memory in mind and evolve from mission to mission. On the other hand, the short-term GP is utilized to adapt with online time-varying disturbances and (sudden) changes in dynamics. It is worth mentioning that, the short-term GP learns around on the predicted mean value of the long-term GP, and the posterior is recursively updated during the online learning phase. Hence, the unknown dynamics $\bm{\Delta}(\tilde{\bm{x}})$ can be rewritten as a sum of two independent GPs: \vspace{-1mm}
\begin{subequations}
\begin{align}
\label{25}
\bm{\Delta}(\tilde{\bm{x}}) &= \bm{\Delta}_l(\tilde{\bm{x}})+\bm{\Delta}_s(\tilde{\bm{x}}) \\
\bm{\Delta}_l(\tilde{\bm{x}}) &\sim\mathcal{GP}(\bm{\mu}_l(\bm{\tilde{x}}),{k}(\bm{\tilde{x}},\bm{\tilde{x}}')),\\ \bm{\Delta}_s(\tilde{\bm{x}})&\sim\mathcal{GP}(\bm{0},{v}(\bm{\tilde{x}},\bm{\tilde{x}}'))
\end{align}\vskip -1mm
\end{subequations}
\noindent where the subscript $l$ and $s$ represent "long-term" and "short-term", respectively. {${v}(\bm{\tilde{x}},\bm{\tilde{x}}')$ denotes the covariance function for the short-term GP. For the simplicity of notation, the prior mean function is set as zero in the following derivation. Furthermore, the mean function $\bm{\mu}_l(\bm{\tilde{x}})$ for the long-term GP is obtained from the offline training phase}. Then, considering the training set $\mathcal{D}$, the prior distribution for $\bm{\Delta}_l$ and $\bm{\Delta}_s$ are given by $P(\bm{\Delta}_l)=\mathcal{N}(\bm{\Delta}_l|\bm{\mu}_l,\bm{K}_{N})$ and $P(\bm{\Delta}_s)=\mathcal{N}(\bm{\Delta}_s|\bm{0},\bm{V}_{N})$. Moreover, the model likelihood is denoted as 
$P(\bm{Y}|\bm{\Delta}_l,\bm{\Delta}_s) = \mathcal{N}(\bm{Y}|\bm{\Delta}_l+\bm{\Delta}_s,\sigma_{\epsilon}^2\bm{I}_{N})$
{where $\bm{V}_{N}$ represents the kernel matrix with $[\bm{V}]_{ij}\!=\!v(\bm{\tilde{x}}_i,\bm{\tilde{x}}_j)$.}
To deal with the large data sets, we introduce two sets of pseudo points for the DGP, i.e., $\bm{u}_l=\bm{\Delta}_l(\bm{z}_l)$ and $\bm{u}_s=\bm{\Delta}_s(\bm{z}_s)$ where $\bm{z}_{(\cdot)}\!=\![\bm{z}_{(\cdot)1},...,\bm{z}_{(\cdot)M}]$ denote the pseudo inputs for the long/short-term GP, respectively. Subsequently, one can obtain the following conditional distributions:
$P(\bm{\Delta}_l|\bm{u}_l) = \mathcal{N}(\bm{\Delta}_l|\bm{K}_{NM}\bm{K}_{M}^{-1}\bm{u}_l, \bm{K}_{N}- \bm{Q}_{N})$ 
and $P(\bm{\Delta}_s|\bm{u}_s) = \mathcal{N}(\bm{\Delta}_s|\bm{V}_{NM}\bm{V}_{M}^{-1}\bm{u}_s, \bm{V}_{N}- \bm{O}_{N})$, where $\bm{O}_{N}=\bm{V}_{NM}\bm{V}_{M}^{-1}\bm{V}_{MN}$. 
The marginal likelihood gives $\log P(\bm{Y})=\log \int P(\bm{Y}|\bm{\Delta}_l,\bm{\Delta}_s)P(\bm{\Delta}_l|\bm{u}_l)P(\bm{\Delta}_s|\bm{u}_s)P(\bm{u}_l)P(\bm{u}_s)\mathrm{d}\bm{\Delta}_l\\\mathrm{d}\bm{\Delta}_s\mathrm{d}\bm{u}_l\mathrm{d}\bm{u}_s$. Next, two variational distributions $q(\bm{u}_l)=\mathcal{N}(\bm{u}_l|\bm{m}_{ul},\bm{S}_{ul})$ and $q(\bm{u}_s)=\mathcal{N}(\bm{u}_s|\bm{m}_{us},\bm{S}_{us})$ are introduced to approximate the posterior: 
$P(\bm{\Delta}_l,\bm{\Delta}_s,\bm{u}_l,\bm{u}_s|\bm{Y})\approx P(\bm{\Delta}_l|\bm{u}_l)P(\bm{\Delta}_s|\bm{u}_s)q(\bm{u}_l)q(\bm{u}_s)$. 
The marginalized distribution of $q(\bm{\Delta}_l) = \int P(\bm{\Delta}_l|\bm{u}_l)q(\bm{u}_l)\mathrm{d}\bm{u}_l$ and  $q(\bm{\Delta}_s) = \int P(\bm{\Delta}_s|\bm{u}_s)q(\bm{u}_s)\mathrm{d}\bm{u}_s$ can be simply computed as 
{\setlength\abovedisplayskip{3pt}
 \setlength\belowdisplayskip{3pt}
\begin{subequations}
\label{31}
\begin{align}
\!\!\!\!q(\bm{\Delta}_l) &\!=\! \mathcal{N}(\bm{K}_{NM}\bm{K}_{M}^{-1}\bm{m}_{ul},\bm{K}_{N}\!+\!\hat{\bm{Q}}_{N})\!\triangleq\!\mathcal{N}(\bm{m}_{l},\bm{S}_{l})\\[0.5mm]
\!\!\!\!q(\bm{\Delta}_s) &\!=\! \mathcal{N}(\bm{V}_{NM}\bm{V}_{M}^{-1}\bm{m}_{us},\bm{V}_{N}\!+\!\hat{\bm{O}}_{N})\!\triangleq\!\mathcal{N}(\bm{m}_{s},\bm{S}_{s})
\end{align}
\end{subequations}
\noindent where $\hat{\bm{Q}}_{N}= \bm{K}_{NM}\bm{K}_{M}^{-1}(\bm{S}_{ul}-\bm{K}_{M})\bm{K}_{M}^{-1}\bm{K}_{MN}$, and  $\hat{\bm{O}}_{N}= \bm{V}_{NM}\bm{V}_{M}^{-1}(\bm{S}_{us}-\bm{V}_{M})\bm{V}_{M}^{-1}\bm{V}_{MN}$. It is worth mentioning that the posterior for long-term GP $q(\bm{\Delta}_l)$ is obtained after the offline training phase and it is fixed during the inference. Furthermore, the posterior for the short-term GP $q(\bm{u}_s)$ is updated with~\eqref{24}.
Then we can obtain the lower bound of $\log P(\bm{Y})$ for DGP as follows:
\begin{multline}
\label{32}
\mathcal{L}(q) \triangleq \int q\left(\bm{\Delta}_l,\bm{\Delta}_s,\bm{u}_l,\bm{u}_s\right) \log \frac{P\left(\bm{Y}, \bm{\Delta}_l,\bm{\Delta}_s,\bm{u}_l,\bm{u}_s\right)}{q\left(\bm{\Delta}_l,\bm{\Delta}_s,\bm{u}_l,\bm{u}_s\right)}\\
\quad\quad\mathrm{d}\bm{\Delta}_l\mathrm{d}\bm{\Delta}_s\mathrm{d}\bm{u}_l\mathrm{d}\bm{u}_s\\
=\int q(\bm{u}_l)\int P(\bm{\Delta}_s|\bm{u}_s)q(\bm{u}_s)\int P(\bm{\Delta}_l|\bm{u}_l)\\
\quad \log P(\bm{Y}|\bm{\Delta}_l,\bm{\Delta}_s)\mathrm{d}\bm{\Delta}_l\mathrm{d}\bm{\Delta}_s\mathrm{d}\bm{u}_s\mathrm{d}\bm{u}_l\\
-\mathcal{KL}(q\left(\bm{u}_l\right)||P\left(\bm{u}_l\right))-\mathcal{KL}(q\left(\bm{u}_s\right)||P\left(\bm{u}_s\right))
\end{multline}
Due to the fact that the posterior of the long-term GP is fixed during the online learning phase, we can equate the variational parameters of $q(\bm{u}_l)$ to~\eqref{18}:  $\bm{m}_{ul}=\bm{m}_{u}$ and $\bm{S}_{ul}=\bm{S}_{u}$. Next we will compute the analytical form of~\eqref{32} to get the optimal variational parameters for $q(\bm{u}_s)$. The inner integral for the first term of~\eqref{32} can be derived as:
{\setlength\abovedisplayskip{3pt}
\be 
\label{33}
\begin{aligned}
&\int P(\bm{\Delta}_l|\bm{u}_l) \log P(\bm{Y}|\bm{\Delta}_l,\bm{\Delta}_s)\mathrm{d}\bm{\Delta}_l \\
&=\int\!\mathcal{N}(\bm{\Delta}_l|\bm{K}_{NM}\bm{K}_{M}^{-1}\bm{u}_l, \bm{K}_{N}\!-\!\bm{Q}_{N})
\log P(\bm{Y}|\bm{\Delta}_l,\bm{\Delta}_s)\mathrm{d}\bm{\Delta}_l\\
&=\log \mathcal{N}(\bm{Y}|\bm{K}_{NM}\bm{K}_{M}^{-1}\bm{u}_l+\bm{\Delta}_s,\sigma_{\epsilon}^2\bm{I}_{N})\!-\!\frac{1}{2\sigma_{\epsilon}^2}\mathrm{tr}(\bm{K}_{N}\!-\!\bm{Q}_{N})
\end{aligned}
\ee
}
By merging inner integral into the second integral in~\eqref{32}:
\begin{multline}
\label{34}
\int \!P(\bm{\Delta}_s|\bm{u}_s)q(\bm{u}_s)\!\int\! P(\bm{\Delta}_l|\bm{u}_l)\log P(\bm{Y}|\bm{\Delta}_l,\bm{\Delta}_s)\mathrm{d}\bm{\Delta}_s\mathrm{d}\bm{u}_s\\
=\int \!q(\bm{\Delta}_s)\log \mathcal{N}(\bm{Y}|\bm{K}_{NM}\bm{K}_{M}^{-1}\bm{u}_l\!+\!\bm{\Delta}_s,\sigma_{\epsilon}^2\bm{I}_{N})\mathrm{d}\bm{\Delta}_s\\
\quad-\frac{1}{2\sigma_{\epsilon}^2}\mathrm{tr}(\bm{K}_{N}- \bm{Q}_{N})\\
=\log \mathcal{N}(\bm{Y}|\bm{K}_{NM}\bm{K}_{M}^{-1}\bm{u}_l+\bm{m}_s,\sigma_{\epsilon}^2\bm{I}_{N})\\
\quad-\frac{1}{2\sigma_{\epsilon}^2}\mathrm{tr}(\bm{K}_{N}- \bm{Q}_{N})-\frac{1}{2\sigma_{\epsilon}^2}\mathrm{tr}(\bm{S}_s).
\end{multline}
}
Substituting~\eqref{34} back to~\eqref{32}, one has: \vspace{-2mm}
\begin{align}
\notag \mathcal{L}(q)&=\int q(\bm{u}_l)\log \mathcal{N}(\bm{Y}|\bm{K}_{NM}\bm{K}_{M}^{-1}\bm{u}_l+\bm{m}_s,\sigma_{\epsilon}^2\bm{I}_{N})\mathrm{d}\bm{u}_l\\ \notag
&\quad-\mathcal{KL}(q\left(\bm{u}_l\right)||P\left(\bm{u}_l\right))-\mathcal{KL}(q\left(\bm{u}_s\right)||P\left(\bm{u}_s\right))\\ 
&\quad-\frac{1}{2\sigma_{\epsilon}^2}\mathrm{tr}(\bm{K}_{N}- \bm{Q}_{N})-\frac{1}{2\sigma_{\epsilon}^2}\mathrm{tr}(\bm{S}_s)\label{35} \\ \notag
&\leq \log\int\mathcal{N}(\bm{Y}|\bm{K}_{NM}\bm{K}_{M}^{-1}\bm{u}_l+\bm{m}_s,\sigma_{\epsilon}^2\bm{I}_{N})P\left(\bm{u}_l\right)\mathrm{d}\bm{u}_l\\[-1mm]\notag
&-\mathcal{KL}(q\left(\bm{u}_s\right)\!||P\left(\bm{u}_s\right))\!-\!\frac{1}{2\sigma_{\epsilon}^2}\mathrm{tr}(\bm{K}_{N}\!-\!\bm{Q}_{N})\!-\!\frac{1}{2\sigma_{\epsilon}^2}\mathrm{tr}(\bm{S}_s)
\end{align}
 
Using Gaussian identities, the integral in the log-term can be further simplified as:
$\int\mathcal{N}(\bm{Y}|\bm{K}_{NM}\bm{K}_{M}^{-1}\bm{u}_l+\bm{m}_s,\sigma_{\epsilon}^2\bm{I}_{N})P\left(\bm{u}_l\right)\mathrm{d}\bm{u}_l
=\mathcal{N}(\bm{Y}|\bm{m}_s,\bm{Q}_{N}+\sigma_{\epsilon}^2\bm{I}_{N})$. 
Furthermore, the KL-divergence between two Gaussians is analytical, that is: $\mathcal{KL}(q\left(\bm{u}_s\right)||P\left(\bm{u}_s\right))=\frac{1}{2}\log|\bm{V}_M|+\frac{1}{2}\bm{m}_{us}^{\top}\bm{V}_M^{-1}\bm{m}_{us}+\frac{1}{2}\mathrm{tr}(\bm{S}_{us}\bm{V}_M^{-1})-\frac{1}{2}M-\frac{1}{2}\log|\bm{S}_{us}|$. Consequently, one gets the analytical form of $\mathcal{L}(q)$:
{\setlength\abovedisplayskip{3pt}
 \setlength\belowdisplayskip{5pt}
\begin{align}
\notag \mathcal{L}(q)&\!=\!\log\mathcal{N}(\bm{Y}|\bm{m}_s,\bm{Q}_{N}\!+\!\sigma_{\epsilon}^2\bm{I}_{N})\!-\!\frac{1}{2\sigma_{\epsilon}^2}\mathrm{tr}(\bm{K}_{N}\!-\!\bm{Q}_{N})\\ \label{37}
&\quad\!-\!\frac{1}{2\sigma_{\epsilon}^2}\mathrm{tr}(\bm{S}_s)\!-\!\frac{1}{2}\log|\bm{V}_M|\!-\!\frac{1}{2}\bm{m}_{us}^{\top}\bm{V}_M^{-1}\bm{m}_{us}\\ \notag
&\quad\!-\!\frac{1}{2}\mathrm{tr}(\bm{S}_{us}\bm{V}_M^{-1})\!+\!\frac{1}{2}M\!+\!\frac{1}{2}\log|\bm{S}_{us}|.
\end{align}}
Taking the derivative of $\mathcal{L}(q)$ with respect to the variational parameters of $q(\bm{u}_s)$ and set them as zeros, we can obtain the optimal $q^{*}(\bm{u}_s)$ with mean and variance satisfy:
{\setlength\abovedisplayskip{3pt}
 \setlength\belowdisplayskip{-3pt}
\be
\label{38}
\begin{aligned}
&\bm{m}_{us}=\bm{V}_{M}(\bm{V}_{M}+\bm{V}_{MN}\tilde{\bm{Q}}_{N}^{-1}\bm{V}_{NM})^{-1}\bm{V}_{MN}\tilde{\bm{Q}}_{N}^{-1}\bm{Y}\\
&\bm{S}_{us}= \bm{V}_{M}\left(\bm{V}_{M}+\sigma_{\epsilon}^{-2} \bm{V}_{M N} \bm{V}_{N M}\right)^{-1} \bm{V}_{M}
\end{aligned}
\ee
}

\noindent where $\tilde{\bm{Q}}_{N}=\bm{Q}_{N}+\sigma_{\epsilon}^2\bm{I}_{N}$. {Combining~\eqref{31} with the learned variational parameters, one can derive the predictive distribution of both of the GPs:}
{\setlength\abovedisplayskip{3pt}
 \setlength\belowdisplayskip{-1pt}
\be
\label{39}
\begin{aligned}
q(\bm{\Delta}_l^{*}) &= \int P(\bm{\Delta}_l^{*}|\bm{u}_l)q^{*}(\bm{u}_l)\mathrm{d}\bm{u}_l^{*}\triangleq\mathcal{N}(\bm{\mu}_{l}^{*},\bm{\Sigma}_{l}^{*})\\[-1mm]
q(\bm{\Delta}_s^{*}) &= \int P(\bm{\Delta}_s^{*}|\bm{u}_s)q^{*}(\bm{u}_s)\mathrm{d}\bm{u}_s^{*}\triangleq\mathcal{N}(\bm{\mu}_{s}^{*},\bm{\Sigma}_{s}^{*})
\end{aligned}
\ee}

\noindent with $\bm{\mu}_{l}^{*}=\bm{K}_{*M}\bm{K}_{M}^{-1}\bm{m}_{ul}$, $\bm{\mu}_{s}^{*}=\bm{V}_{*M}\bm{V}_{M}^{-1}\bm{m}_{us}$, $\bm{\Sigma}_{l}^{*}=k_{**}-\bm{K}_{*M}\bm{K}_{M}^{-1}\bm{K}_{M*}+\bm{K}_{*M}\bm{K}_{M}^{-1}\bm{S}_{ul}\bm{K}_{M}^{-1}\bm{K}_{M*}$, and $\bm{\Sigma}_{s}^{*}=v_{**}-\bm{V}_{*M}\bm{V}_{M}^{-1}\bm{V}_{M*}+\bm{V}_{*M}\bm{V}_{M}^{-1}\bm{S}_{us}\bm{V}_{M}^{-1}\bm{V}_{M*}$.

Finally, the predictive distribution of the latent function $\bm{\Delta}$ at a new test point $\tilde{\bm{x}}^{*}$ is given by
$q(\bm{\Delta}^{*}) = \int P(\bm{\Delta}^{*}|\bm{\Delta}_l^{*},\bm{\Delta}_s^{*})q(\bm{\Delta}_l^{*})q(\bm{\Delta}_s^{*})\mathrm{d}\bm{\Delta}_l^{*}\mathrm{d}\bm{\Delta}_s^{*}$ which leads to the following predictive mean and variance of the DGP:
\begin{align}
&\bm{\mu}_{\Delta}(\bm{\tilde{x}}^{*})=\bm{K}_{*M}\bm{K}_{M}^{-1}\bm{m}_{ul}\!+\!\bm{V}_{*M}\bm{V}_{M}^{-1}\bm{m}_{us} \label{41}\\ \notag
&\bm{\Sigma}_{\Delta}(\bm{\tilde{x}}^{*})=k_{**}\!+\!v_{**}\!-\!\bm{K}_{*M}\bm{K}_{M}^{-1}\bm{K}_{M*}\!-\!\bm{V}_{*M}\bm{V}_{M}^{-1}\bm{V}_{M*}\\ \notag
&\quad+\bm{K}_{*M}\bm{K}_{M}^{-1}\bm{S}_{ul}\bm{K}_{M}^{-1}\bm{K}_{M*}\!+\!\bm{V}_{*M}\bm{V}_{M}^{-1}\bm{S}_{us}\bm{V}_{M}^{-1}\bm{V}_{M*}
\end{align}
The predictive distribution $q(\bm{\Delta}^{*})$ of the latent function $\bm{\Delta}$ can be used for multi-step prediction over a given horizon as it is done in the MPC scheme in the next section. Furthermore, to update the predictive mean and variance of the short-term GP, one can first recursively update the posterior distribution $q(\bm{u}_s)$  with~\eqref{24} using online data points, then marginalize the posterior on $\bm{u}_s$ as in~\eqref{31}, leading to $q(\bm{\Delta}^{*})$.
\vspace{-3mm}
\section{GP-MPC for quadcopters}
 \YL{To design a stabilizing tracking controller for the system~\eqref{6} subject to the robust satisfaction of state and control input constraints $\mathcal{X}\subseteq \mathbb{R}^{12}$ and $\mathcal{U}\subseteq \mathbb{R}^{4}$, }the following constrained optimal control problem is formulated as:\vspace{-1.5mm}
\be
\label{42}
\begin{aligned}
\min_{\bm{u}(k)}~ &J=\mathbb{E}\left[l_T(\bm{x}_{H|k},\bm{r}_{H|k})\!+\!\!\sum_{i=0}^{H-1}l(\bm{x}_{i|k},\bm{u}_{i|k},\bm{r}_{i|k})\right]\\
\rm{s.t.}~  &\bm{x}_{0|k}=\bm{x}(k), \bm{u}_{0|k}=\bm{u}(k),\\
&\bm{x}_{i+1|k}=\bm{A}\bm{x}_{i|k}+\bm{B}\bm{u}_{i|k}+\bm{B}_d\bm{\Delta}(\bm{x}_{i|k},\bm{u}_{i|k}),\\
&\bm{x}_{i|k}\in \mathcal{X}, ~\bm{u}_{i|k}\in \mathcal{U},
\end{aligned}
\ee
\noindent where $H$ is the prediction horizon, and $\bm{r}(\cdot)$ represents the desired trajectory to be tracked. $l(\cdot)$ and $l_T(\cdot)$ are the stage and terminal cost functions, which are defined in a quadratic form with weight matrices $\bm{Q}$, $\bm{Q}_T$ and $\bm{R}$: $l(\cdot)\triangleq\|\bm{x}_{i|k}-\bm{r}_{i|k}\|_{\bm{Q}}^{2}+\|\bm{u}_{i|k}\|_{\bm{R}}^{2}$, and $l_T(\cdot)\triangleq\|\bm{x}_{H|k}-\bm{r}_{H|k}\|_{\bm{Q}_T}^{2}$.

 In~\eqref{42}, the model uncertainty $\bm{\Delta}(\bm{x}_{i|k},\bm{u}_{i|k})=\mathcal{N}(\bm{\Delta}|\bm{\mu}_{\Delta}^{k},\bm{\Sigma}_{\Delta}^{k})$ is a stochastic term which is inferred by the DGP structure. This directly leads to a probabilistic prediction model $P(\bm{x}_{i+1|k})=\mathcal{N}(\bm{\mu}_{x}^{i+1},\bm{\Sigma}_{x}^{i+1})$. Assuming that the predictive state is time independent, then one can obtain the mean and variance of the prediction model as:
\be
\label{43}
\ba{l}
\bm{\mu}_{x}^{i+1}=\bm{A}\bm{\mu}_{x}^{i}+\bm{B}\bm{u}_{i|k}+\bm{B}_d\bm{\mu}_{\Delta}^{i},\\[1mm]
\bm{\Sigma}_{x}^{i+1}=\bm{A}\bm{\Sigma}_{x}^{i}\bm{A}^{\top}+\bm{B}_d\bm{\Sigma}_{\Delta}^{i}\bm{B}_d^{\top}
\ea
\ee

It can be observed from~\eqref{43} that the input for GP $\bm{\Delta}(\bm{x}_{k},\bm{u}_{k})$ becomes a random variable during the state propagation, which results in a non-Gaussian posterior and is intractable to be computed analytically. We adopt the exact moment matching approach\cite{quinonero2003prediction,deisenroth2010efficient} to fit the posterior optimally with the first and second moments. One can use other moment matching methods to obtain the cost distribution as well. This allows us to express the stochastic cost function as a deterministic form:
\be
\label{44}
\begin{aligned}
&J=(\bm{\mu}^{H|k}_{x}\!-\!\bm{r}_{H|k})^{\top}\bm{Q}_{T}(\bm{\mu}^{H|k}_{x}\!-\!\bm{r}_{H|k})+\mathrm{tr}(\bm{Q}_{T}\bm{\Sigma}_{x}^{H|k})+\\
&\sum_{i=0}^{H-1}\!((\bm{\mu}^{i|k}_{x}\!-\!\bm{r}_{i|k})^{\top}\bm{Q}(\bm{\mu}^{i|k}_{x}\!-\!\bm{r}_{i|k})\!+\!\bm{u}_{i|k}^{\top}\bm{R}\bm{u}_{i|k}\!+\!\mathrm{tr}(\bm{Q}\bm{\Sigma}_{x}^{i|k}))
\end{aligned}
\ee

Next, we discuss the satisfaction of the constraints on the state and the control input. Consider linear state constraints $\bm{H}\bm{x}_{i|k}\leq \bm{h}$ with $\bm{H}_j\bm{x}_{i|k}\leq{h}_j$, $j=1,...,m$, $\bm{H}\in\mathbb{R}^{m\times 12}$ and $\bm{h}\in\mathbb{R}^{m}$. As $\bm{x}_{i|k}$ is a random variable with mean $\bm{\mu}_{x}^{i}$ and variance $\bm{\Sigma}_{x}^{i}$,  we intend to satisfy the constraints in probabilistic sense, i.e. by the chance constraint: $P(\bm{H}\bm{x}_{i|k}\leq \bm{h})\geq\gamma$. The quantile function $\bm{\varpi}(\gamma)$, which can be seen as the inverse CDF, is utilized to convert the above chance constraints into deterministic ones. Denoting $\bm{z}=\bm{H}\bm{x}_{i|k}-\bm{h}$, then we have $\bm{z}\sim\mathcal{N}(\bm{H}\bm{\mu}^{i|k}_{x}-\bm{h},\bm{H}\bm{\Sigma}_{x}^{i|k}\bm{H}^{\top})$. Thus, according to the Galois inequalities, the chance constraint is equivalent to $\bm{\varpi}(\gamma)\leq 0$ and finally leads to
\[
\bm{H}\bm{\mu}^{i|k}_{x}\leq \bm{h}-\bm{\varpi}(\gamma)\sqrt{\bm{H}\bm{\Sigma}_{x}^{i|k}\bm{H}^{\top}}.
\] On the other hand, the control input constraints $\mathcal{U}$ is simply defined as
$\bm{u}_{\rm{min}}\leq\bm{u}_{i|k}\leq\bm{u}_{\rm{max}}$.
Finally, the GP-MPC problem in this paper is formulated as:
\be
{\label{48}
\begin{aligned}
\min_{\bm{u}(k)}~&~\eqref{44}\\
{\rm{s.t.}} ~  \bm{x}_{0|k}&=\bm{x}(k), \bm{u}_{0|k}=\bm{u}(k)\\
\bm{\mu}_{x}^{i+1}&=\bm{A}\bm{\mu}_{x}^{i}+\bm{B}\bm{u}_{i}+\bm{B}_d\bm{\mu}_{\Delta}^{i},\\
\bm{\Sigma}_{x}^{i+1}&=\bm{A}\bm{\Sigma}_{x}^{i}\bm{A}^{\top}+\bm{B}_d\bm{\Sigma}_{\Delta}^{i}\bm{B}_d^{\top}\\
\bm{H}\bm{\mu}^{i|k}_{x}&\leq \bm{h}-\bm{\varpi}(\gamma)\sqrt{\bm{H}\bm{\Sigma}_{x}^{i|k}\bm{H}^{\top}}\\
\bm{u}_{\rm{min}}&\leq\bm{u}_{i|k}\leq\bm{u}_{\rm{max}} 
\end{aligned}}
\ee
\YL{Because the stochastic cost has been transformed into a deterministic form, most nonlinear optimization algorithms can be utilized to solve the problem.}

\section{SIMULATION STUDY}

In this section, numerical simulation results are provided to illustrate the effectiveness of the proposed dual GP based MPC strategy for quadcopters. The parameters of the quadcopter model are considered according to \cite{pounds2007design}. \YL{The quadcopter is forced to track a helix-like trajectory in 3D space, which is described by: $x(t)=2\sin(t)$,  $y(t)=2\cos(t)$, and $z(t)=\{2,3,4\}$ for every third of total operation time $T$.} The sampling time $T_s$ is 0.1 s. The discrete-time nonlinear system dynamics of the quadcopter are linearized and decomposed into two LTI systems, i.e., the outer-loop translational subsystem and the inner-loop rotational subsystem. To simplify the controller design and speed up the simulation, we adopt the PD controller for the rotational subsystem and use the MPC for the translational subsystem. \YL{It is worth mentioning that, despite the linearizing effect feedback control in the rotational subsystem driven by PD controller, the nonlinearities and additional dynamics experienced there affect the translational system, which interprets the necessity of a GP compensation.} For the MPC scheme, the prediction horizon is chosen as  $H=5$, and the weight matrices in the cost function are: $\bm{Q}=\mathrm{diag}(1~ 1~ 20 ~1~ 1 ~20)$, and $\bm{R}=\mathrm{diag}(1~ 1 ~1)$. 

First, to construct the training data set, the quadcopter is enabled to track a random reference trajectory and explore the space as much as possible. In general, the data can also be collected by flying the quadcopter manually. We use the linear MPC as the exciting input for the quadcopter for 50s, and then collect the data with 20Hz, resulting in a training data set with size $N = 1000$.  Furthermore, we employ a constant wind with speed $[1~3~-2]^{\top}$ m/s in frame $\mathcal{I}$ which occurs as a heading dependent uncertainty in dynamics~\eqref{1} during the offline training phase, and a time-varying wind with speed $[1~3~-2]^{\top}+\sin(\frac{\pi}{10}(t-10))*[-2~-3~3]^{\top}$ m/s, which appears as a time-varying orientation dependent uncertainty into the system at $t=10$ s. Then, a SVGP is trained offline with $M=20$ pseudo inputs, which gives the long-term GP for the first iteration. At the same time, the short-term GP is initialized as a zero-mean GP.  

\begin{figure}[t]
 \captionsetup{font={small}}
  \includegraphics[width= 3in]{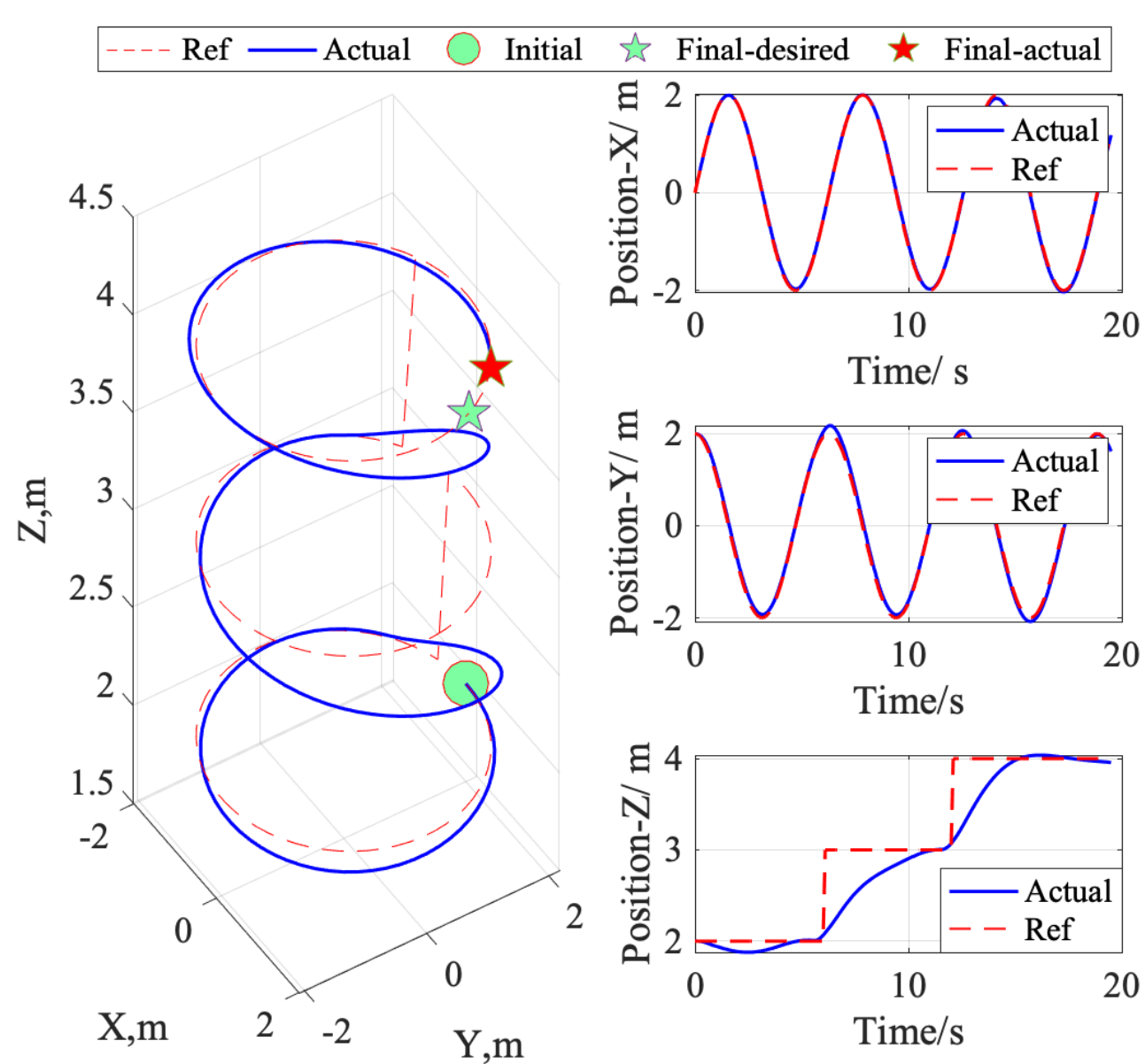}
  \caption{Trajectory tracking with DGP-MPC (after the 3rd iteration).} 
  \label{fig:4} \vspace{-3mm}
\end{figure}

\begin{figure}[t] 
\begin{center}
 \includegraphics[width= 3.1in]{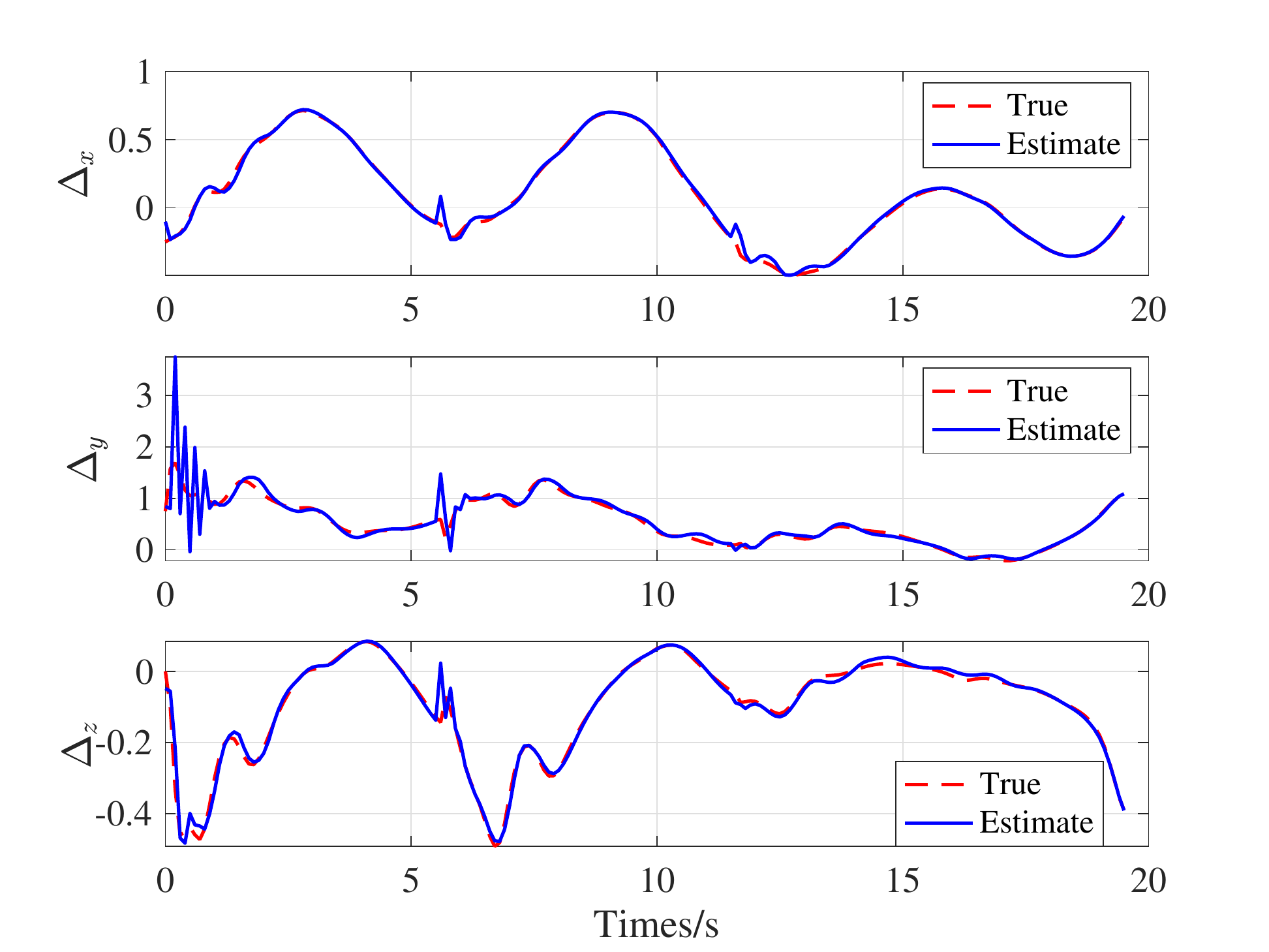}
 \end{center} \vspace{-3mm}
  \captionsetup{font={small}}
  \caption{DGP estimation results.} 
  \label{fig:5} \vspace{-3mm}
\end{figure}

\begin{figure}[t]
\begin{center}
  \includegraphics[width= 3in]{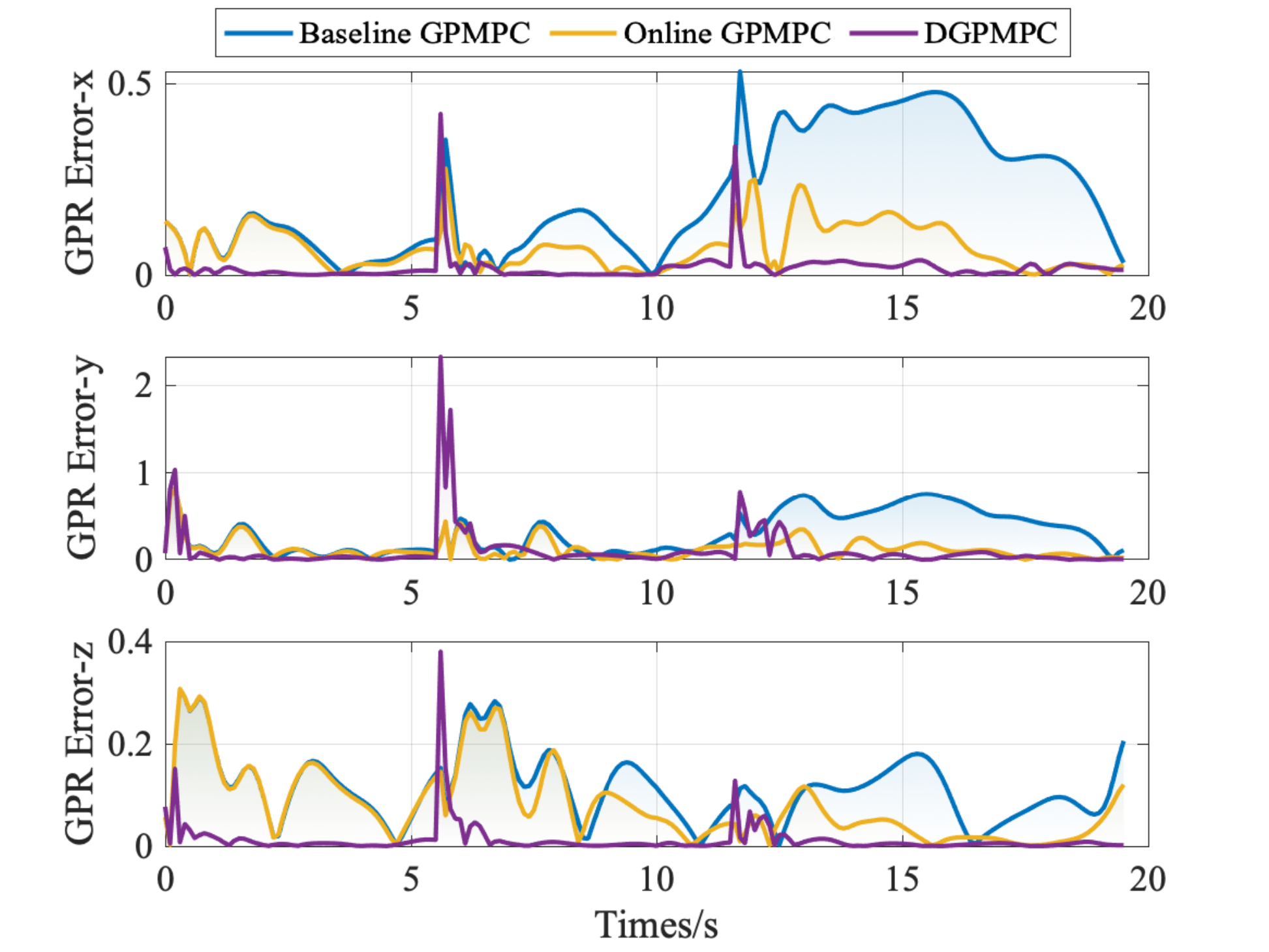}
  \end{center} \vspace{-3mm}
  \captionsetup{font={small}}
  \caption{GP regression error of three methods.} 
  \label{fig:6}
\end{figure}

To show the advantage of the proposed DGP-MPC strategy, a baseline GP-MPC scheme without online recursive update is employed in the simulations. The simulation results are shown in Figs.~\ref{fig:4} - \ref{fig:6}. 
The trajectory tracking result of the proposed DGP-MPC after the 3rd iteration is depicted in Fig.~\ref{fig:4}. The tracking performance is adequate though the constant wind from 0s - 10s  turns into time-varying wind at 10s. In addition, the DGP estimation results and error comparison among the baseline GPMPC, online recursive GPMPC and the proposed DGPMPC are shown in Figs.~\ref{fig:5} - \ref{fig:6}.  It is obvious that both the online GP and DGP can recursively update with the new measurement data after 10s, and DGP shows better performance on estimating the uncertain model owing to the "memory" data. Furthermore, to quantify the performance and tracking errors of the proposed controller and baseline controller, the average cost $\bar{J}$ within the prediction horizon and the mean square error (MSE) are recorded in Table. 1.

\begin{table}[t]
\centering
\caption{Quantitative evaluation of the two controllers}
\begin{tabular}{llllll}
\hline
\multicolumn{2}{c}{\multirow{2}{*}{Controller}} & \multicolumn{1}{c}{\multirow{2}{*}{$\bar{J}$}} & \multicolumn{3}{c}{MSE}  \\ \cline{4-6} 
\multicolumn{2}{c}{}                            & \multicolumn{1}{c}{}                                          & $x$      & $y$      & $z$      \\ \hline
\multicolumn{2}{l}{Baseline GP-MPC}                    & 35.9730                                                               &  0.0020     &  0.0053      & 0.0914       \\
\multirow{2}{*}{DGP-MPC}       & 1st iter.       & 36.4046                                                       & 0.0023 & 0.0062 & 0.0928 \\
                              & 3rd iter.       & 33.3544                                                       & 0.0011 & 0.0035 & 0.0825 \\ \hline
\end{tabular}
\vspace{-3.5mm}
\end{table}

\YL{One can see that comparing with the baseline controller, the performance of the proposed DGP-MPC strategy in the third iteration has been improved. The reason why the $\bar{J}$ and MSE for baseline GP-MPC are smaller than DGP-MPC in the first iteration is that the estimation of $\bm{\Delta}$ needs to converge to the true value during the first seconds of online recursion in DGP-MPC. Moreover, comparing the $\bar{J}$ and MSE within the iterations of DGP-MPC, one can observe that after incorporating new data into the long-term GP from mission to mission, the tracking error decreases obviously. The reason is that the long-term GP gradually captures more and more model uncertainties after the batch training.}

\vspace{-1mm}
\section{CONCLUSIONS}
This paper presents a novel Dual Gaussian Process based model predictive control for quadcopter trajectory tracking with model uncertainties and time-varying external disturbances. The knowledge of learned model is kept in the long-term GP while the short-term GP performs recursive online adaption to compensate the unlearned uncertainties during control operation. The DGP structure is not limited to MPC schemes, but can be also extended to any GP-based controllers which require both "memory" and "learning". Future work will focus on the real-time implementation of the proposed strategy on Crazyflie nano quadcopters with VICON positioning system.

\vspace{-3mm}
\bibliographystyle{IEEEtran}
\bibliography{IEEEfull,reference}

\end{document}